\documentclass[10pt,a4paper,onecolumn]{article}
\usepackage{marginnote}
\usepackage{graphicx}
\usepackage{xcolor}
\usepackage{authblk,etoolbox}
\usepackage{titlesec}
\usepackage{calc}
\usepackage{tikz}
\usepackage{hyperref}
\hypersetup{colorlinks,breaklinks,
            urlcolor=[rgb]{0.0, 0.5, 1.0},
            linkcolor=[rgb]{0.0, 0.5, 1.0}}
\usepackage{caption}
\usepackage{tcolorbox}
\usepackage{amssymb,amsmath}
\usepackage{ifxetex,ifluatex}
\usepackage{seqsplit}
\usepackage{xstring}

\usepackage{float}
\let\origfigure\figure
\let\endorigfigure\endfigure
\renewenvironment{figure}[1][2] {
    \expandafter\origfigure\expandafter[H]
} {
    \endorigfigure
}

\usepackage{fixltx2e} % provides \textsubscript
\usepackage[
  backend=biber,
]{biblatex}
\bibliography{paper.bib}

\let\textttOrig=\texttt
\def\texttt#1{\expandafter\textttOrig{\seqsplit{#1}}}
\renewcommand{\seqinsert}{\ifmmode
  \allowbreak
  \else\penalty6000\hspace{0pt plus 0.02em}\fi}

\makeatletter
\let\href@Orig=\href
\def\href@Urllike#1#2{\href@Orig{#1}{\begingroup
    \def\Url@String{#2}\Url@FormatString
    \endgroup}}
\def\href@Notdoi#1#2{\def\tempa{#1}\def\tempb{#2}%
  \ifx\tempa\tempb\relax\href@Urllike{#1}{#2}\else
  \href@Orig{#1}{#2}\fi}
\def\href#1#2{%
  \IfBeginWith{#1}{https://doi.org}%
  {\href@Urllike{#1}{#2}}{\href@Notdoi{#1}{#2}}}
\makeatother

\usepackage[top=3.5cm, bottom=3cm, right=1.5cm, left=1.0cm,
            headheight=2.2cm, reversemp, includemp, marginparwidth=4.5cm]{geometry}

\titleformat{\section}
  {\normalfont\sffamily\Large\bfseries}
  {}{0pt}{}
\titleformat{\subsection}
  {\normalfont\sffamily\large\bfseries}
  {}{0pt}{}
\titleformat{\subsubsection}
  {\normalfont\sffamily\bfseries}
  {}{0pt}{}
\titleformat*{\paragraph}
  {\sffamily\normalsize}

\usepackage{fancyhdr}
\makeatletter
\let\ps@plain\ps@fancy
\fancyheadoffset[L]{4.5cm}
\fancyfootoffset[L]{4.5cm}

\definecolor{linky}{rgb}{0.0, 0.5, 1.0}

\newtcolorbox{repobox}
   {colback=red, colframe=red!75!black,
     boxrule=0.5pt, arc=2pt, left=6pt, right=6pt, top=3pt, bottom=3pt}

\newcommand{\ExternalLink}{%
   \tikz[x=1.2ex, y=1.2ex, baseline=-0.05ex]{%
       \begin{scope}[x=1ex, y=1ex]
           \clip (-0.1,-0.1)
               --++ (-0, 1.2)
               --++ (0.6, 0)
               --++ (0, -0.6)
               --++ (0.6, 0)
               --++ (0, -1);
           \path[draw,
               line width = 0.5,
               rounded corners=0.5]
               (0,0) rectangle (1,1);
       \end{scope}
       \path[draw, line width = 0.5] (0.5, 0.5)
           -- (1, 1);
       \path[draw, line width = 0.5] (0.6, 1)
           -- (1, 1) -- (1, 0.6);
       }
   }

\patchcmd{\@maketitle}{center}{flushleft}{}{}
\patchcmd{\@maketitle}{center}{flushleft}{}{}
\patchcmd{\@maketitle}{\LARGE}{\LARGE\sffamily}{}{}
\def\maketitle{{%
  
  \AB@maketitle}}
\makeatletter
\renewcommand\AB@affilsepx{ \protect\Affilfont}
\renewcommand\AB@affilnote[1]{{\bfseries #1}\hspace{3pt}}
\renewcommand{\affil}[2][]%
   {\newaffiltrue\let\AB@blk@and\AB@pand
      \if\relax#1\relax\def\AB@note{\AB@thenote}\else\def\AB@note{#1}%
        \setcounter{Maxaffil}{0}\fi
        \begingroup
        \let\href=\href@Orig
        \let\texttt=\textttOrig
        \let\protect\@unexpandable@protect
        \def\thanks{\protect\thanks}\def\footnote{\protect\footnote}%
        \@temptokena=\expandafter{\AB@authors}%
        {\def\\{\protect\\\protect\Affilfont}\xdef\AB@temp{#2}}%
         \xdef\AB@authors{\the\@temptokena\AB@las\AB@au@str
         \protect\\[\affilsep]\protect\Affilfont\AB@temp}%
         \gdef\AB@las{}\gdef\AB@au@str{}%
        {\def\\{, \ignorespaces}\xdef\AB@temp{#2}}%
        \@temptokena=\expandafter{\AB@affillist}%
        \xdef\AB@affillist{\the\@temptokena \AB@affilsep
          \AB@affilnote{\AB@note}\protect\Affilfont\AB@temp}%
      \endgroup
       \let\AB@affilsep\AB@affilsepx
}
\makeatother

\renewcommand\Affilfont{\sffamily\small\mdseries}
\ifnum 0\ifxetex 1\fi\ifluatex 1\fi=0 % if pdftex
  \usepackage[T1]{fontenc}
  \usepackage[utf8]{inputenc}

\else % if luatex or xelatex
  \ifxetex
    \usepackage{mathspec}
  \else
    \usepackage{fontspec}
  \fi
  \defaultfontfeatures{Ligatures=TeX,Scale=MatchLowercase}

\fi
\IfFileExists{upquote.sty}{\usepackage{upquote}}{}
\IfFileExists{microtype.sty}{%
\usepackage{microtype}
\UseMicrotypeSet[protrusion]{basicmath} % disable protrusion for tt fonts
}{}

\usepackage{hyperref}
\hypersetup{unicode=true,
            pdftitle={The Pencil Code, a modular MPI code for partial differential equations and particles: multipurpose and multiuser-maintained},
            pdfborder={0 0 0},
            breaklinks=true}
\let\addcontentslineOrig=\addcontentsline
\def\addcontentsline#1#2#3{\bgroup
  \let\texttt=\textttOrig\addcontentslineOrig{#1}{#2}{#3}\egroup}
\let\markbothOrig\markboth
\def\markboth#1#2{\bgroup
  \let\texttt=\textttOrig\markbothOrig{#1}{#2}\egroup}
\let\markrightOrig\markright
\def\markright#1{\bgroup
  \let\texttt=\textttOrig\markrightOrig{#1}\egroup}

\usepackage{graphicx,grffile}
\makeatletter
\def\maxwidth{\ifdim\Gin@nat@width>\linewidth\linewidth\else\Gin@nat@width\fi}
\def\maxheight{\ifdim\Gin@nat@height>\textheight\textheight\else\Gin@nat@height\fi}
\makeatother
\setkeys{Gin}{width=\maxwidth,height=\maxheight,keepaspectratio}
\IfFileExists{parskip.sty}{%
\usepackage{parskip}
}{% else
\setlength{\parindent}{0pt}
\setlength{\parskip}{6pt plus 2pt minus 1pt}
}
\providecommand{\tightlist}{%
  \setlength{\itemsep}{0pt}\setlength{\parskip}{0pt}}
\ifx\paragraph\undefined\else
\let\oldparagraph\paragraph
\renewcommand{\paragraph}[1]{\oldparagraph{#1}\mbox{}}
\fi
\ifx\subparagraph\undefined\else
\let\oldsubparagraph\subparagraph
\renewcommand{\subparagraph}[1]{\oldsubparagraph{#1}\mbox{}}
\fi

\title{The Pencil Code, a modular MPI code for partial differential equations
and particles: multipurpose and multiuser-maintained}

        \author[1]{The Pencil Code Collaboration}
          \author[1, 2, 3]{Axel Brandenburg}
          \author[4]{Anders Johansen}
          \author[5]{Philippe A. Bourdin}
          \author[6]{Wolfgang Dobler}
          \author[7]{Wladimir Lyra}
          \author[8]{Matthias Rheinhardt}
          \author[9]{Sven Bingert}
          \author[10, 11, 1]{Nils Erland L. Haugen}
          \author[12]{Antony Mee}
          \author[8, 13]{Frederick Gent}
          \author[14]{Natalia Babkovskaia}
          \author[15]{Chao-Chin Yang}
          \author[16]{Tobias Heinemann}
          \author[17]{Boris Dintrans}
          \author[1]{Dhrubaditya Mitra}
          \author[18]{Simon Candelaresi}
          \author[19]{Jörn Warnecke}
          \author[20]{Petri J. Käpylä}
          \author[14]{Andreas Schreiber}
          \author[21]{Piyali Chatterjee}
          \author[8, 19]{Maarit J. Käpylä}
          \author[1]{Xiang-Yu Li}
          \author[10, 11]{Jonas Krüger}
          \author[11]{Jørgen R. Aarnes}
          \author[13]{Graeme R. Sarson}
          \author[22]{Jeffrey S. Oishi}
          \author[23]{Jennifer Schober}
          \author[24]{Raphaël Plasson}
          \author[1]{Christer Sandin}
          \author[11, 25]{Ewa Karchniwy}
          \author[13, 26]{Luiz Felippe S. Rodrigues}
          \author[27]{Alexander Hubbard}
          \author[28]{Gustavo Guerrero}
          \author[13]{Andrew Snodin}
          \author[1]{Illa R. Losada}
          \author[8]{Johannes Pekkilä}
          \author[29]{Chengeng Qian}
    
      \affil[1]{Nordita, KTH Royal Institute of Technology and Stockholm University,
Sweden}
      \affil[2]{Department of Astronomy, Stockholm University, Sweden}
      \affil[3]{McWilliams Center for Cosmology \& Department of Physics, Carnegie
Mellon University, PA, USA}
      \affil[4]{GLOBE Institute, University of Copenhagen, Denmark}
      \affil[5]{Space Research Institute, Graz, Austria}
      \affil[6]{Bruker, Potsdam, Germany}
      \affil[7]{New Mexico State University, Department of Astronomy, Las Cruces, NM,
USA}
      \affil[8]{Astroinformatics, Department of Computer Science, Aalto University,
Finland}
      \affil[9]{Gesellschaft für wissenschaftliche Datenverarbeitung mbH Göttingen,
Germany}
      \affil[10]{SINTEF Energy Research, Trondheim, Norway}
      \affil[11]{Norwegian University of Science and Technology, Norway}
      \affil[12]{Bank of America Merrill Lynch, London, UK}
      \affil[13]{School of Mathematics, Statistics and Physics, Newcastle University, UK}
      \affil[14]{No current affiliation}
      \affil[15]{University of Nevada, Las Vegas, USA}
      \affil[16]{Niels Bohr International Academy, Denmark}
      \affil[17]{CINES, Montpellier, France}
      \affil[18]{School of Mathematics and Statistics, University of Glasgow, UK}
      \affil[19]{Max Planck Institute for Solar System Research, Germany}
      \affil[20]{Institute for Astrophysics, University of Göttinge, Germany}
      \affil[21]{Indian Institute of Astrophysics, Bengaluru, India}
      \affil[22]{Department of Physics \& Astronomy, Bates College, ME, USA}
      \affil[23]{Laboratoire d'Astrophysique, EPFL, Sauverny, Switzerland}
      \affil[24]{Avignon Université, France}
      \affil[25]{Institute of Thermal Technology, Silesian University of Technology,
Poland}
      \affil[26]{Radboud University, Netherlands}
      \affil[27]{Department of Astrophysics, American Museum of Natural History, NY, USA}
      \affil[28]{Physics Department, Universidade Federal de Minas Gerais, Belo
Horizonte, Brazil}
      \affil[29]{State Key Laboratory of Explosion Science and Technology, Beijing
Institute of Technology, China}
  \date{\vspace{-5ex}}

\begin{document}
\maketitle

\marginpar{
  \sffamily\small

  {\bfseries DOI:} \href{https://doi.org/}{\color{linky}{}}

  \vspace{2mm}

  {\bfseries Software}
  \begin{itemize}
    \setlength\itemsep{0em}
    \item \href{}{\color{linky}{Review}} \ExternalLink
    \item \href{https://github.com/pencil-code/pencil-code}{\color{linky}{Repository}} \ExternalLink
    \item \href{http://dx.doi.org/10.5281/zenodo.3961647}{\color{linky}{Archive}} \ExternalLink
  \end{itemize}

  \vspace{2mm}

  {\bfseries Submitted:} 17 September 2020\\
  {\bfseries Published:} 

  \vspace{2mm}
  {\bfseries License}\\
  Authors of papers retain copyright and release the work under a Creative Commons Attribution 4.0 International License (\href{https://creativecommons.org/licenses/by/4.0/}{\color{linky}{CC BY 4.0}}).
}

\hypertarget{summary}{%
\section{Summary}\label{summary}}

The Pencil Code is a highly modular physics-oriented simulation code
that can be adapted to a wide range of applications. It is primarily
designed to solve partial differential equations (PDEs) of compressible
hydrodynamics and has lots of add-ons ranging from astrophysical
magnetohydrodynamics (MHD) (A. Brandenburg and Dobler 2010) to
meteorological cloud microphysics (Li et al. 2017) and engineering
applications in combustion (Babkovskaia, Haugen, and Brandenburg 2011).
Nevertheless, the framework is general and can also be applied to
situations not related to hydrodynamics or even PDEs, for example when
just the message passing interface or input/output strategies of the
code are to be used. The code can also evolve Lagrangian (inertial and
noninertial) particles, their coagulation and condensation, as well as
their interaction with the fluid. A related module has also been adapted
to perform ray tracing and to solve the eikonal equation.

The code is being used for Cartesian, cylindrical, and spherical
geometries, but further extensions are possible. One can choose between
different time stepping schemes and different spatial derivative
operators. High-order first and second derivatives are used to deal with
weakly compressible turbulent flows. There are also different diffusion
operators to allow for both direct numerical simulations (DNS) and
various types of large-eddy simulations (LES).

\hypertarget{high-level-functionality}{%
\section{High-level functionality}\label{high-level-functionality}}

An idea about the range of available modules can be obtained by
inspecting the examples under pencil-code/samples/. Those are low
resolution versions related to applications published in the literature.
Some of the run directories of actual production runs are published
through Zenodo. Below a list of method papers that describe the various
applications and tests:

\begin{itemize}
\tightlist
\item
  Coagulation and condensation in turbulence (Johansen et al. 2008; Li
  et al. 2017),
\item
  Radiative transfer (Heinemann et al. 2006; Barekat and Brandenburg
  2014; A. Brandenburg and Das 2020),
\item
  Chiral magnetic effect in relativistic plasmas (Schober et al. 2018),
\item
  Primordial gravitational waves (Roper Pol et al. 2020),
\item
  Modeling homochirality at the origin of life (Brandenburg and
  Multamäki 2004; Brandenburg 2019),
\item
  Modeling of patterned photochemical systems (Emond et al. 2012),
\item
  Gaseous combustion and detonation (Babkovskaia, Haugen, and
  Brandenburg 2011; Zhang et al. 2020; Krüger, Haugen, and Løvås 2017),
\item
  Burning particles, resolved or unresolved (Qian et al. 2020),
\item
  Flows around immersed solid objects (Aarnes, Haugen, and Andersson
  2019; Aarnes et al. 2020; Haugen and Kragset 2010),
\item
  Test-field method for turbulent MHD transport (Rheinhardt and
  Brandenburg 2010; A. Brandenburg et al. 2010; Warnecke et al. 2018),
\item
  Mean-field MHD (Kemel et al. 2013; Jabbari et al. 2013),
\item
  Spherical shell dynamos and convection (Mitra et al. 2009; Käpylä et
  al. 2020),
\item
  Boris correction for coronal physics (Chatterjee 2020),
\item
  Thermal instability and mixing (Yang and Krumholz 2012),
\item
  Implicit solver for temperature (Gastine and Dintrans 2008),
\item
  Dust-gas dynamics with mutual drag interaction (Youdin and Johansen
  2007; Yang and Johansen 2016).
\end{itemize}

\hypertarget{statement-of-need-and-purpose-of-software}{%
\section{Statement of need and purpose of
software}\label{statement-of-need-and-purpose-of-software}}

The code is an easily adaptable tool for solving both standard MHD
equations as well as others, such as the test-field equations.
Significant amounts of runtime diagnostics as well as Python and IDL
libraries for post-processing are available.

Among the currently 83 developers with check-in permission, there are
currently 18 owners who can give others check-in permission. Of the
developers, 34 have done more than 34 commits. Users have access to the
latest development version and can ask to join the circle of developers
by contacting one of the owners.

Every revision on GitHub is verified on 9 tests on travis-ci.com. The
current version is also automatically being tested on 59 hourly tests
and on 79 daily tests. Continuous progress on the code is driven by the
research of individual developers.

Further developments and interactions between developers and users are
being promoted through annual user meetings since 2004 and a newsletters
since 2020. Since 2016, a steering committee of five elected owners
reviews the progress and can take decisions of general concern to the
Pencil Code community.

\hypertarget{ongoing-research-using-the-pencil-code}{%
\section{Ongoing research using the Pencil
Code}\label{ongoing-research-using-the-pencil-code}}

Current research includes topics from stellar physics, interstellar and
intercluster medium, the early universe, as well as from meteorology and
engineering: small-scale dynamos and reconnection; primordial magnetic
fields and decaying turbulence; gravitational waves from turbulent
sources; planet formation and inertial particles; accretion discs and
shear flows; coronal heating and coronal mass ejections; helical
dynamos, helical turbulence, and catastrophic quenching;
helioseismology; strongly stratified MHD turbulence and negative
effective magnetic pressure instability; convection in Cartesian
domains; global convection and dynamo simulations; turbulent transport
and test-field methods; hydrodynamic and MHD instabilities and
turbulence; chiral MHD; turbulent gaseous and solid combustion, particle
clustering and deposition on solid walls, front propagation, radiation
\& ionization. As of July 2020, 564 papers have been published that
acknowledge use of the Pencil Code (A. Brandenburg 2020).

\hypertarget{key-references}{%
\section{Key references}\label{key-references}}

The Pencil Code is unique in two ways: the high level of flexibility and
modularity, and the way it is organized (open source, distributed
ownership, openness of development version).

Other software addressing related needs include: Athena, CO5BOLD, ENZO,
MuRAM, NIRVANA, Stagger, ZEUS, and several other LES codes. There are
also several other engineering DNS codes such as Sandia-3-Dimensional
(S3D), a high-order compressible code, optimized for combustion, which
is not open source, however. In addition, there are frameworks like
Dedalus or Cactus, which allow one to program the equations in symbolic
form.

Some recent research areas that made use of the Pencil Code, as
evidenced by the aforementioned document listing all those papers (A.
Brandenburg 2020), include:

\begin{itemize}
\tightlist
\item
  Flows around immersed solid objects (Haugen and Kragset 2010),
\item
  Particle clustering in supersonic and subsonic turbulence (Mattsson et
  al. 2019; Karchniwy, Klimanek, and Haugen 2019),
\item
  Cloud microphysics (Li et al. 2017),
\item
  Planet and planetesimal formation (Johansen et al. 2007; Oishi, Mac
  Low, and Menou 2007; Lyra et al. 2009),
\item
  Global simulations of debris disks (Lyra and Kuchner 2013),
\item
  Stratified shearing box simulations, also with dust (Oishi and Mac Low
  2011; Schreiber and Klahr 2018; Yang, Mac Low, and Johansen 2018),
\item
  Supernova-driven turbulence (Gent et al. 2013),
\item
  Solar dynamo and sunspots (Brandenburg 2005; Heinemann et al. 2007),
\item
  Solar corona above active regions (Bingert and Peter 2011; Bourdin,
  Bingert, and Peter 2013; Chatterjee, Hansteen, and Carlsson 2016),
\item
  Fully convective star in a box (Dobler, Stix, and Brandenburg 2006),
\item
  Dynamo wave in spherical shell convection (Käpylä, Mantere, and
  Brandenburg 2012; Warnecke et al. 2014),
\item
  Convection with Kramers opacity law (Käpylä et al. 2017, 2020; Käpylä
  2019),
\item
  MHD turbulence and cascades (Haugen, Brandenburg, and Dobler 2004),
\item
  Turbulent diffusivity quenching with test fields (Brandenburg et al.
  2008; Karak et al. 2014).
\end{itemize}

\hypertarget{acknowledgements}{%
\section{Acknowledgements}\label{acknowledgements}}

We acknowledge contributions from all submitters and their supporting
funding agencies. In particular, we mention the ERC Advanced Grant on
Astrophysical Dynamos (No 227952), the Swedish Research Council, grants
2012-5797, 2013-03992, 2017-03865, and 2019-04234, the National Science
Foundation under the grant AAG-1615100, the FRINATEK grant 231444 under
the Research Council of Norway, SeRC, the grant ``Bottlenecks for
particle growth in turbulent aerosols'' from the Knut and Alice
Wallenberg Foundation, Dnr.~KAW 2014.0048, the ReSoLVE Centre of
Excellence (grant number 307411), the research project `Gaspro',
financed by the Research Council of Norway (267916), the European
Research Council (ERC) under the European Union's Horizon 2020 research
and innovation programme (Project UniSDyn, grant agreement n:o 818665),
and the Deutsche Forschungsgemeinschaft (DFG) Heisenberg programme grant
KA 4825/2-1.

\hypertarget{references}{%
\section*{References}\label{references}}
\addcontentsline{toc}{section}{References}

\hypertarget{refs}{}
\leavevmode\hypertarget{ref-2019IJCFD.33.43A}{}%
Aarnes, Jørgen R., Nils E. L. Haugen, and Helge I. Andersson. 2019.
``High-order overset grid method for detecting particle impaction on a
cylinder in a cross flow.'' \emph{Int. J. Comput. Fluid Dynam.} 33
(February): 43--58. \url{https://doi.org/10.1080/10618562.2019.1593385}.

\leavevmode\hypertarget{ref-2020GApFD.114.35A}{}%
Aarnes, Jørgen R., Tai Jin, Chaoli Mao, Nils E. L. Haugen, Kun Luo, and
Helge I. Andersson. 2020. ``Treatment of solid objects in the Pencil
Code using an immersed boundary method and overset grids.''
\emph{Geophys. Astrophys. Fluid Dynam.} 114 (March): 35--57.
\url{https://doi.org/10.1080/03091929.2018.1492720}.

\leavevmode\hypertarget{ref-2011JCoPh.230.1B}{}%
Babkovskaia, N., N. E. L. Haugen, and A. Brandenburg. 2011. ``A
high-order public domain code for direct numerical simulations of
turbulent combustion.'' \emph{J. Comp. Phys.} 230 (January): 1--12.
\url{https://doi.org/10.1016/j.jcp.2010.08.028}.

\leavevmode\hypertarget{ref-2014Aux5cux26A.571A.68B}{}%
Barekat, A., and A. Brandenburg. 2014. ``Near-polytropic stellar
simulations with a radiative surface'' 571 (November): A68.
\url{https://doi.org/10.1051/0004-6361/201322461}.

\leavevmode\hypertarget{ref-2011Aux5cux26A.530A.112B}{}%
Bingert, S., and H. Peter. 2011. ``Intermittent heating in the solar
corona employing a 3D MHD model'' 530 (June): A112.
\url{https://doi.org/10.1051/0004-6361/201016019}.

\leavevmode\hypertarget{ref-2013Aux5cux26A.555A.123B}{}%
Bourdin, P.-A., S. Bingert, and H. Peter. 2013. ``Observationally driven
3D magnetohydrodynamics model of the solar corona above an active
region'' 555 (July): A123.
\url{https://doi.org/10.1051/0004-6361/201321185}.

\leavevmode\hypertarget{ref-2005ApJ.625.539B}{}%
Brandenburg, A. 2005. ``The case for a distributed solar dynamo shaped
by near-surface shear'' 625 (May): 539--47.
\url{https://doi.org/10.1086/429584}.

\leavevmode\hypertarget{ref-zenodo.3466444}{}%
---------. 2020. ``Scientific usage of the Pencil Code.'' 2020.
\url{http://doi.org/10.5281/zenodo.3466444}.

\leavevmode\hypertarget{ref-2010PhST.142a4028B}{}%
Brandenburg, A., P. Chatterjee, F. Del Sordo, A. Hubbard, P. J. Käpylä,
and M. Rheinhardt. 2010. ``Turbulent transport in hydromagnetic flows.''
\emph{Phys. Scripta Vol. T} 142 (December): 014028.
\url{https://doi.org/10.1088/0031-8949/2010/T142/014028}.

\leavevmode\hypertarget{ref-2020GApFD.114.162B}{}%
Brandenburg, A., and U. Das. 2020. ``The time step constraint in
radiation hydrodynamics.'' \emph{Geophys. Astrophys. Fluid Dynam.} 114
(March): 162--95. \url{https://doi.org/10.1080/03091929.2019.1676894}.

\leavevmode\hypertarget{ref-2010ascl.soft10060B}{}%
Brandenburg, A., and W. Dobler. 2010. ``Pencil Code: Finite-difference
Code for Compressible Hydrodynamic Flows.''
\url{https://doi.org/10.5281/zenodo.2315093}.

\leavevmode\hypertarget{ref-2004IJAsB.3.209B}{}%
Brandenburg, A., and T. Multamäki. 2004. ``How long can left and right
handed life forms coexist?'' \emph{Int. J. Astrobiology} 3 (July):
209--19. \url{https://doi.org/10.1017/S1473550404001983}.

\leavevmode\hypertarget{ref-2008ApJ.676.740B}{}%
Brandenburg, A., K.-H. Rädler, M. Rheinhardt, and P. J. Käpylä. 2008.
``Magnetic diffusivity tensor and dynamo effects in rotating and
shearing turbulence'' 676 (March): 740--51.
\url{https://doi.org/10.1086/527373}.

\leavevmode\hypertarget{ref-2019OLEB.49.49B}{}%
Brandenburg, Axel. 2019. ``The limited roles of autocatalysis and
enantiomeric cross-inhibition in achieving homochirality in dilute
systems.'' \emph{Origins of Life and Evolution of the Biosphere} 49
(June): 49--60. \url{https://doi.org/10.1007/s11084-019-09579-4}.

\leavevmode\hypertarget{ref-2020GApFD.114.213C}{}%
Chatterjee, Piyali. 2020. ``Testing Alfvén wave propagation in a
`realistic' set-up of the solar atmosphere.'' \emph{Geophys. Astrophys.
Fluid Dynam.} 114 (March): 213--34.
\url{https://doi.org/10.1080/03091929.2019.1672676}.

\leavevmode\hypertarget{ref-2016PhRvL.116j1101C}{}%
Chatterjee, Piyali, Viggo Hansteen, and Mats Carlsson. 2016. ``Modeling
repeatedly flaring \(\delta\) sunspots'' 116 (March): 101101.
\url{https://doi.org/10.1103/PhysRevLett.116.101101}.

\leavevmode\hypertarget{ref-2006ApJ.638.336D}{}%
Dobler, W., M. Stix, and A. Brandenburg. 2006. ``Magnetic field
generation in fully convective rotating spheres'' 638 (February):
336--47. \url{https://doi.org/10.1086/498634}.

\leavevmode\hypertarget{ref-2012ChemEurJ}{}%
Emond, Matthieu, Thomas Le Saux, Jean-Francois Allemand, Philippe
Pelupessy, Raphaël Plasson, and Ludovic Jullien. 2012. ``Energy
Propagation Through a Protometabolism Leading to the Local Emergence of
Singular Stationary Concentration Profiles.'' \emph{Chem. Eur. J.} 18
(45): 14375--83. \url{https://doi.org/10.1002/chem.201201974}.

\leavevmode\hypertarget{ref-2008Aux5cux26A.484.29G}{}%
Gastine, T., and B. Dintrans. 2008. ``Direct numerical simulations of
the \(\kappa\)-mechanism. I. Radial modes in the purely radiative case''
484 (June): 29--42. \url{https://doi.org/10.1051/0004-6361:20078936}.

\leavevmode\hypertarget{ref-2013MNRAS.432.1396G}{}%
Gent, F. A., A. Shukurov, A. Fletcher, G. R. Sarson, and M. J. Mantere.
2013. ``The supernova-regulated ISM - I. The multiphase structure'' 432
(June): 1396--1423. \url{https://doi.org/10.1093/mnras/stt560}.

\leavevmode\hypertarget{ref-2004PhRvE.70a6308H}{}%
Haugen, N. E., A. Brandenburg, and W. Dobler. 2004. ``Simulations of
nonhelical hydromagnetic turbulence'' 70 (July): 016308.
\url{https://doi.org/10.1103/PhysRevE.70.016308}.

\leavevmode\hypertarget{ref-2010JFM.661a239}{}%
Haugen, Nils Erland L., and Steinar Kragset. 2010. ``Particle Impaction
on a Cylinder in a Crossflow as Function of Stokes and Reynolds
Numbers.'' \emph{Journal of Fluid Mechanics} 661: 239--61.
\url{https://doi.org/10.1017/S0022112010002946}.

\leavevmode\hypertarget{ref-2006Aux5cux26A.448.731H}{}%
Heinemann, T., W. Dobler, Å. Nordlund, and A. Brandenburg. 2006.
``Radiative transfer in decomposed domains'' 448 (March): 731--37.
\url{https://doi.org/10.1051/0004-6361:20053120}.

\leavevmode\hypertarget{ref-2007ApJ.669.1390H}{}%
Heinemann, T., Å. Nordlund, G. B. Scharmer, and H. C. Spruit. 2007.
``MHD simulations of penumbra fine structure'' 669 (November): 1390--4.
\url{https://doi.org/10.1086/520827}.

\leavevmode\hypertarget{ref-2013Aux5cux26A.556A.106J}{}%
Jabbari, S., A. Brandenburg, N. Kleeorin, D. Mitra, and I. Rogachevskii.
2013. ``Surface flux concentrations in a spherical \(\alpha\)\(^{2}\)
dynamo'' 556 (August): A106.
\url{https://doi.org/10.1051/0004-6361/201321353}.

\leavevmode\hypertarget{ref-2008Aux5cux26A.486.597J}{}%
Johansen, A., F. Brauer, C. Dullemond, H. Klahr, and T. Henning. 2008.
``A coagulation-fragmentation model for the turbulent growth and
destruction of preplanetesimals'' 486 (August): 597--611.
\url{https://doi.org/10.1051/0004-6361:20079232}.

\leavevmode\hypertarget{ref-2007Natur.448.1022J}{}%
Johansen, A., J. S. Oishi, M.-M. Mac Low, H. Klahr, T. Henning, and A.
Youdin. 2007. ``Rapid planetesimal formation in turbulent circumstellar
disks'' 448 (August): 1022--5.
\url{https://doi.org/10.1038/nature06086}.

\leavevmode\hypertarget{ref-2017ApJ.845.23K}{}%
Käpylä, Petri J., Matthias Rheinhardt, Axel Brandenburg, Rainer Arlt,
Maarit J. Käpylä, Andreas Lagg, Nigul Olspert, and Jörn Warnecke. 2017.
``Extended subadiabatic layer in simulations of overshooting
convection'' 845 (August): L23.
\url{https://doi.org/10.3847/2041-8213/aa83ab}.

\leavevmode\hypertarget{ref-2019Aux5cux26A.631.122K}{}%
Käpylä, P. J. 2019. ``Overshooting in simulations of compressible
convection'' 631 (November): A122.
\url{https://doi.org/10.1051/0004-6361/201834921}.

\leavevmode\hypertarget{ref-2020GApFD.114.8K}{}%
Käpylä, P. J., F. A. Gent, N. Olspert, M. J. Käpylä, and A. Brandenburg.
2020. ``Sensitivity to luminosity, centrifugal force, and boundary
conditions in spherical shell convection.'' \emph{Geophys. Astrophys.
Fluid Dynam.} 114 (March): 8--34.
\url{https://doi.org/10.1080/03091929.2019.1571586}.

\leavevmode\hypertarget{ref-2012ApJ.755L.22K}{}%
Käpylä, P. J., M. J. Mantere, and A. Brandenburg. 2012. ``Cyclic
magnetic activity due to turbulent convection in spherical wedge
geometry'' 755 (August): L22.
\url{https://doi.org/10.1088/2041-8205/755/1/L22}.

\leavevmode\hypertarget{ref-2014ApJ.795.16K}{}%
Karak, B. B., M. Rheinhardt, A. Brandenburg, P. J. Käpylä, and M. J.
Käpylä. 2014. ``Quenching and anisotropy of hydromagnetic turbulent
transport'' 795 (November): 16.
\url{https://doi.org/10.1088/0004-637X/795/1/16}.

\leavevmode\hypertarget{ref-Karchniwy_etal_2019}{}%
Karchniwy, Ewa, Adam Klimanek, and Nils Erland L. Haugen. 2019. ``The
effect of turbulence on mass transfer rates between inertial particles
and fluid for polydisperse particle size distributions.'' \emph{J. Fluid
Mech.} 874: 1147--68. \url{https://doi.org/10.1017/jfm.2019.493}.

\leavevmode\hypertarget{ref-2013SoPh.287.293K}{}%
Kemel, K., A. Brandenburg, N. Kleeorin, D. Mitra, and I. Rogachevskii.
2013. ``Active region formation through the negative effective magnetic
pressure instability'' 287 (October): 293--313.
\url{https://doi.org/10.1007/s11207-012-0031-8}.

\leavevmode\hypertarget{ref-2017CNF.185a160}{}%
Krüger, Jonas, Nils Erland L. Haugen, and Terese Løvås. 2017.
``Correlation effects between turbulence and the conversion rate of
pulverized char particles.'' \emph{Combustion and Flame} 185: 160--72.
\url{https://doi.org/10.1016/j.combustflame.2017.07.008}.

\leavevmode\hypertarget{ref-2017JAMES.9.1116L}{}%
Li, Xiang-Yu, A. Brandenburg, N. E. L. Haugen, and G. Svensson. 2017.
``Eulerian and Lagrangian approaches to multidimensional condensation
and collection.'' \emph{J. Adv. Model. Earth Systems} 9 (June):
1116--37. \url{https://doi.org/10.1002/2017MS000930}.

\leavevmode\hypertarget{ref-2009Aux5cux26A.497.869L}{}%
Lyra, W., A. Johansen, A. Zsom, H. Klahr, and N. Piskunov. 2009.
``Planet formation bursts at the borders of the dead zone in 2D
numerical simulations of circumstellar disks'' 497 (April): 869--88.
\url{https://doi.org/10.1051/0004-6361/200811265}.

\leavevmode\hypertarget{ref-2013Natur.499.184L}{}%
Lyra, W., and M. Kuchner. 2013. ``Formation of sharp eccentric rings in
debris disks with gas but without planets'' 499 (July): 184--87.
\url{https://doi.org/10.1038/nature12281}.

\leavevmode\hypertarget{ref-2019MNRAS.483.5623M}{}%
Mattsson, Lars, Akshay Bhatnagar, Fred A. Gent, and Beatriz Villarroel.
2019. ``Clustering and dynamic decoupling of dust grains in turbulent
molecular clouds'' 483 (March): 5623--41.
\url{https://doi.org/10.1093/mnras/sty3369}.

\leavevmode\hypertarget{ref-2009ApJ.697.923M}{}%
Mitra, D., R. Tavakol, A. Brandenburg, and D. Moss. 2009. ``Turbulent
dynamos in spherical shell segments of varying geometrical extent'' 697
(May): 923--33. \url{https://doi.org/10.1088/0004-637X/697/1/923}.

\leavevmode\hypertarget{ref-2011ApJ.740.18O}{}%
Oishi, J. S., and M.-M. Mac Low. 2011. ``Magnetorotational turbulence
transports angular momentum in stratified disks with low magnetic
Prandtl number but magnetic Reynolds number above a critical value'' 740
(October): 18. \url{https://doi.org/10.1088/0004-637X/740/1/18}.

\leavevmode\hypertarget{ref-2007ApJ.670.805O}{}%
Oishi, J. S., M.-M. Mac Low, and K. Menou. 2007. ``Turbulent torques on
protoplanets in a dead zone'' 670 (November): 805--19.
\url{https://doi.org/10.1086/521781}.

\leavevmode\hypertarget{ref-2020GApFD.114.58Q}{}%
Qian, Chengeng, Cheng Wang, JianNan Liu, Axel Brand enburg, Nils E. L.
Haugen, and Mikhael A. Liberman. 2020. ``Convergence properties of
detonation simulations.'' \emph{Geophys. Astrophys. Fluid Dynam.} 114
(March): 58--76. \url{https://doi.org/10.1080/03091929.2019.1668382}.

\leavevmode\hypertarget{ref-2010Aux5cux26A.520A.28R}{}%
Rheinhardt, M., and A. Brandenburg. 2010. ``Test-field method for
mean-field coefficients with MHD background'' 520 (September): A28.
\url{https://doi.org/10.1051/0004-6361/201014700}.

\leavevmode\hypertarget{ref-2020GApFD.114.130R}{}%
Roper Pol, Alberto, Axel Brandenburg, Tina Kahniashvili, Arthur
Kosowsky, and Sayan Mandal. 2020. ``The timestep constraint in solving
the gravitational wave equations sourced by hydromagnetic turbulence.''
\emph{Geophys. Astrophys. Fluid Dynam.} 114 (March): 130--61.
\url{https://doi.org/10.1080/03091929.2019.1653460}.

\leavevmode\hypertarget{ref-2018ApJ.858.124S}{}%
Schober, Jennifer, Igor Rogachevskii, Axel Brandenburg, Alexey Boyarsky,
Jürg Fröhlich, Oleg Ruchayskiy, and Nathan Kleeorin. 2018. ``Laminar and
turbulent dynamos in chiral magnetohydrodynamics. II. simulations'' 858
(May): 124. \url{https://doi.org/10.3847/1538-4357/aaba75}.

\leavevmode\hypertarget{ref-2018ApJ.861.47S}{}%
Schreiber, Andreas, and Hubert Klahr. 2018. ``Azimuthal and vertical
streaming instability at high dust-to-gas ratios and on the scales of
planetesimal formation'' 861 (July): 47.
\url{https://doi.org/10.3847/1538-4357/aac3d4}.

\leavevmode\hypertarget{ref-2014ApJ.796L.12W}{}%
Warnecke, J., P. J. Käpylä, M. J. Käpylä, and A. Brandenburg. 2014. ``On
the cause of solar-like equatorward migration in global convective
dynamo simulations'' 796 (November): L12.
\url{https://doi.org/10.1088/2041-8205/796/1/L12}.

\leavevmode\hypertarget{ref-2018Aux5cux26A.609A.51W}{}%
Warnecke, J., M. Rheinhardt, S. Tuomisto, P. J. Käpylä, M. J. Käpylä,
and A. Brandenburg. 2018. ``Turbulent transport coefficients in
spherical wedge dynamo simulations of solar-like stars'' 609 (January):
A51. \url{https://doi.org/10.1051/0004-6361/201628136}.

\leavevmode\hypertarget{ref-2016ApJS.224.39Y}{}%
Yang, C.-C., and A. Johansen. 2016. ``Integration of particle-gas
systems with stiff mutual drag interaction'' 224 (June): 39.
\url{https://doi.org/10.3847/0067-0049/224/2/39}.

\leavevmode\hypertarget{ref-2012ApJ.758.48Y}{}%
Yang, C.-C., and M. Krumholz. 2012. ``Thermal-instability-driven
turbulent mixing in galactic disks. I. Effective mixing of metals'' 758
(October): 48. \url{https://doi.org/10.1088/0004-637X/758/1/48}.

\leavevmode\hypertarget{ref-2018ApJ.868.27Y}{}%
Yang, Chao-Chin, Mordecai-Mark Mac Low, and Anders Johansen. 2018.
``Diffusion and concentration of solids in the dead zone of a
protoplanetary disk'' 868 (November): 27.
\url{https://doi.org/10.3847/1538-4357/aae7d4}.

\leavevmode\hypertarget{ref-2007ApJ.662.613Y}{}%
Youdin, A., and A. Johansen. 2007. ``Protoplanetary Disk Turbulence
Driven by the Streaming Instability: Linear Evolution and Numerical
Methods'' 662 (June): 613--26. \url{https://doi.org/10.1086/516729}.

\leavevmode\hypertarget{ref-Zhang_etal_2020comb}{}%
Zhang, Hancong, Kun Luo, Nils Erland L. Haugen, Chaoli Mao, and Jianren
Fan. 2020. ``Drag force for a burning particle.'' \emph{Comb. Flame}
217: 188--99. \url{https://doi.org/10.1016/j.combustflame.2020.02.016}.

\end{document}